# Barrier-bound States in Flat-band Systems

Jhinhwan Lee

*Department of Physics, Korea Advanced Institute of Science and Technology, Daejeon 34141, Republic of Korea*

**As the fundamental quantum mechanical theory predicts, it is believed that electronic states can be bound only to potential wells and not to potential barriers in any dimension and their energies should be below the background potential. However, with the help of atomic lattice potentials with a flat electron band in momentum space in the background, an atomically thin 1D potential barrier can possess barrier-centered bound states above the background potential. Here we provide a theoretical proof using Green function pole analysis with Dyson equation that shows the existence of barrier-bound-states whose energy is elevated from the flat band energy nearly by the barrier height. The phenomenon is believed to be independently confirmed in a system of Pd monolayer on a flat oxide substrate with Stoner ferromagnetism using spin-polarized STM technique and through comparison with real-space QPI simulations using realistic 2D Pd band structure.**

### Subatomic-scale calculation-lattice-based Dyson equation

Let's consider a simple one-band system with no explicit spin degree of freedom for simplicity. The bare Green function can be written as:

$$G^0(\boldsymbol{k},\omega) = \frac{1}{\omega - E(\boldsymbol{k}) + i\delta(\omega)} \tag{1}$$

where $\delta(\omega)$ reflects the Fermi or the non-Fermi liquid behavior of the quasiparticles and the impurity scattering rates. The details of $\delta(\omega)$ is to be fitted to the energy-dependent spectral broadening, for example by comparing the spectral function $-\frac{1}{\pi}\mathrm{Im}G^0(\boldsymbol{k},\omega)$ with the data obtained in photo-electron spectroscopy.

The Dyson equation to be used in evaluating the full Green function in the presence of the Coulomb potential is

$$G(\boldsymbol{r},\boldsymbol{r}',\omega) = G^0(\boldsymbol{r}-\boldsymbol{r}',\omega) + \sum_{r'',r'''\in\mathcal{L}(A)} G^0(\boldsymbol{r}-\boldsymbol{r}'',\omega)H'(\boldsymbol{r}'',\boldsymbol{r}''')G(\boldsymbol{r}''',\boldsymbol{r}',\omega) \tag{2}$$

where $\mathcal{L}(A)$ is the atomic lattice sites within the calculation area $A$ [1]. Here $A$ will be assumed to be a square area with physical size of $\alpha a \times \alpha a$ ($a$ : atomic lattice constant) for simplicity, irrespective of the atomic lattice symmetry (rectangular or triangular lattices).

The above Equation 2 can be converted to one with the summations over the subatomic-scale calculation lattice sites $\mathbb{N}(A)$

$$G(\boldsymbol{r},\boldsymbol{r}',\omega) = G^0(\boldsymbol{r}-\boldsymbol{r}',\omega) + \frac{\alpha^4}{N^4\sin^2\theta}\sum_{r'',r'''\in\mathbb{N}(A)} G^0(\boldsymbol{r}-\boldsymbol{r}'',\omega)H'(\boldsymbol{r}'',\boldsymbol{r}''')G(\boldsymbol{r}''',\boldsymbol{r}',\omega) \tag{3}$$

Here $\mathbb{N}(A)$ is an $N \times N$ calculation lattice sites over $A$, and $\theta$ is the angle between the unit-cell basis vectors (e.g. $\theta = 60°$ for a regular triangular lattice) [2].

The real-space Green functions are given by

$$G^0(\boldsymbol{r},\omega) = \sum_{\boldsymbol{k}\in\boldsymbol{1^{st}BZ}}^{M} G^0(\boldsymbol{k},\omega)e^{i\boldsymbol{k}\cdot\boldsymbol{r}} \tag{4}$$

where $M$ is the total number of allowed $\boldsymbol{k}$ points contained in the 1st Brilloiun zone [3]. This can be converted to one with summations over $\mathbb{N}_{\boldsymbol{k}}^{\boldsymbol{1}}$ which is the $m$-element 1st BZ subset of $\mathbb{N}_{\boldsymbol{k}}$ in $\boldsymbol{k}$ space.

$$G^0(\boldsymbol{r},\omega) = \beta\sum_{\boldsymbol{k}\in\mathbb{N}_{\boldsymbol{k}}^{1}}^{m} G^0(\boldsymbol{k},\omega)e^{i\boldsymbol{k}\cdot\boldsymbol{r}} \tag{5}$$

The $\beta$ factor is determined by requiring that a constant potential $H_0\delta_{\boldsymbol{r}'',\boldsymbol{r}'''}I$ replacing $H'(\boldsymbol{r}'',\boldsymbol{r}''')$ in the Eq. 3 should lead to a rigid energy shift of the Green function:

$$G(\boldsymbol{r},\boldsymbol{r}',\omega) = G^0(\boldsymbol{r}-\boldsymbol{r}',\omega) + \frac{\alpha^4}{N^4\sin^2\theta}\sum_{r'',r'''\in\mathbb{N}(A)} G^0(\boldsymbol{r}-\boldsymbol{r}'',\omega)H_0\delta_{\boldsymbol{r}'',\boldsymbol{r}'''}IG(\boldsymbol{r}''',\boldsymbol{r}',\omega)$$

$$= G^0(\boldsymbol{r}-\boldsymbol{r}',\omega-H_0) \tag{6}$$

This can be shown that, for $\boldsymbol{r}' = \boldsymbol{0}$ without loss of generality, the assumed equality of the right-hand-sides of Eq. 6 that is to be used to determine $\beta$ is given by

$$\beta\sum_{\boldsymbol{k}\in\mathbb{N}_{\boldsymbol{k}}^{1}}^{m} G_0(\boldsymbol{k},\omega-H_0)e^{i\boldsymbol{k}\cdot\boldsymbol{r}} = \beta\sum_{\boldsymbol{k}\in\mathbb{N}_{\boldsymbol{k}}^{1}}^{m} e^{i\boldsymbol{k}\cdot\boldsymbol{r}}\left[G_0(\boldsymbol{k},\omega) + \beta\frac{m\alpha^4}{N^4\sin^2\theta}H_0G_0(\boldsymbol{k},\omega)G_0(\boldsymbol{k},\omega-H_0)\right] \tag{7}$$

We can equate both sides of the Eq. 7 term by term and have the desired rigid energy shift:

$$G_0(\boldsymbol{k},\omega - H_0) = \frac{G_0(\boldsymbol{k},\omega)}{1-\beta\frac{m\alpha^4}{N^4\sin^2\theta}H_0 G_0(\boldsymbol{k},\omega)} = \frac{1}{G_0(\boldsymbol{k},\omega)^{-1}-\beta\frac{m\alpha^4}{N^4\sin^2\theta}H_0} = G_0\left(\boldsymbol{k},\omega - \beta\frac{m\alpha^4}{N^4\sin^2\theta}H_0\right) \quad (8)$$

Therefore we require $\beta = \frac{N^4\sin^2\theta}{m\alpha^4}$ and

$$G^0(\boldsymbol{r},\omega) = \frac{N^4\sin^2\theta}{m\alpha^4}\sum_{\boldsymbol{k}\in\mathbb{N}_{\boldsymbol{k}}^{1}}^{m} G^0(\boldsymbol{k},\omega)e^{i\boldsymbol{k}\cdot\boldsymbol{r}} \quad (9)$$

Note that the product of the two correction factors in Eqs. 3 and 9 is $1/m$ where $m = N^2\frac{\text{Area of }\mathbb{N}_{\boldsymbol{k}}^{1}}{\text{Area of }\mathbb{N}_{\boldsymbol{k}}}$ is in fact equal to the total number of lattice sites in $A$ ($m = \alpha^2$ for a square lattice and $m = \frac{2\alpha^2}{\sqrt{3}}$ for a regular triangular lattice).

**Proof of existence of barrier-bound states in flat-band systems**

Let's start from the limiting case of a completely flat band, i.e. $E(\boldsymbol{k}) \equiv E_0$. Then $G^0(\boldsymbol{k},\omega) = (\omega - E_0 + i\delta)^{-1}$ and due to the Eq. 9,

$$G^0(\boldsymbol{r},\omega) = \frac{N^4\sin^2\theta}{m\alpha^4}\sum_{\boldsymbol{k}\in\mathbb{N}_{\boldsymbol{k}}^{1}}^{m}\frac{e^{i\boldsymbol{k}\cdot\boldsymbol{r}}}{\omega-E_0+i\delta} = C\frac{A(\boldsymbol{r})}{\omega-E_0+i\delta} \quad (10)$$

where $C \equiv m\beta = \frac{N^4\sin^2\theta}{\alpha^4}$ and $A(\boldsymbol{r}) \equiv \frac{1}{m}\sum_{\boldsymbol{k}\in\mathbb{N}_{\boldsymbol{k}}^{1}}^{m} e^{i\boldsymbol{k}\cdot\boldsymbol{r}}$ is a sharply peaked function near $\boldsymbol{r}$=**0** whose width is about a few times the size of the lattice constant and satisfies $A(\boldsymbol{0}) = 1$.

Then, for a Coulomb potential $H'(\boldsymbol{r},\boldsymbol{r}') = U(\boldsymbol{r})\delta_{\boldsymbol{r},\boldsymbol{r}'}$, the Eq. 3 becomes

$$\bar{G}(\boldsymbol{r},\boldsymbol{r}',\omega) = \bar{G}^0(\boldsymbol{r}-\boldsymbol{r}',\omega) + \sum_{\boldsymbol{r}''\in\mathbb{N}(A)}\bar{G}^0(\boldsymbol{r}-\boldsymbol{r}'',\omega)U(\boldsymbol{r}'')\bar{G}(\boldsymbol{r}'',\boldsymbol{r}',\omega) \quad (11)$$

where $\bar{G} \equiv G/C$ and $\bar{G}^0 \equiv G^0/C$. Here $U(\boldsymbol{r})$ is the perturbing barrier potential in addition to the regular atomic lattice potential, i.e. $U(\boldsymbol{r}) \approx U_0$ for $\boldsymbol{r}$ on the barrier and $U(\boldsymbol{r}) = 0$ for $\boldsymbol{r}$ away from the barrier.

Since $\bar{G}^0(\boldsymbol{r},\omega) \equiv \frac{G^0(\boldsymbol{r},\omega)}{C} = \frac{A(\boldsymbol{r})}{\omega-E_0+i\delta}$, we have

$$\bar{G}(\boldsymbol{r},\boldsymbol{r}',\omega) = \frac{A(\boldsymbol{r}-\boldsymbol{r}')}{\omega-E_0+i\delta} + \sum_{\boldsymbol{r}''\in\mathbb{N}(A)}\frac{A(\boldsymbol{r}-\boldsymbol{r}'')}{\omega-E_0+i\delta}U(\boldsymbol{r}'')\bar{G}(\boldsymbol{r}'',\boldsymbol{r}',\omega) \quad (12)$$

$$(\omega - E_0 + i\delta)\bar{G}(\boldsymbol{r},\boldsymbol{r}',\omega) = A(\boldsymbol{r}-\boldsymbol{r}') + \sum_{\boldsymbol{r}''\in\mathbb{N}(A)}A(\boldsymbol{r}-\boldsymbol{r}'')U(\boldsymbol{r}'')\bar{G}(\boldsymbol{r}'',\boldsymbol{r}',\omega) \quad (13)$$

$$\sum_{\boldsymbol{r}''\in\mathbb{N}(A)}\left[(\omega - E_0 + i\delta)\delta_{\boldsymbol{r},\boldsymbol{r}''} - A(\boldsymbol{r}-\boldsymbol{r}'')U(\boldsymbol{r}'')\right]\bar{G}(\boldsymbol{r}'',\boldsymbol{r}',\omega) = A(\boldsymbol{r}-\boldsymbol{r}') \quad (14)$$

This is a matrix identity in the form of $\bar{F}\bar{G} = A$ in the position basis. The matrices are of size $N^2 \times N^2$ and the elements are given by:

$$\bar{F}_{\boldsymbol{r},\boldsymbol{r}''}(\omega) \equiv (\omega - E_0 + i\delta)\delta_{\boldsymbol{r},\boldsymbol{r}''} - A(\boldsymbol{r} - \boldsymbol{r}'')U(\boldsymbol{r}'') \tag{15}$$

$$\bar{G}_{\boldsymbol{r}'',\boldsymbol{r}'}(\omega) \equiv \bar{G}(\boldsymbol{r}'', \boldsymbol{r}', \omega) \tag{16}$$

$$A_{\boldsymbol{r},\boldsymbol{r}'} \equiv A(\boldsymbol{r} - \boldsymbol{r}') \tag{17}$$

The poles of $\bar{G}$ are approximately (not exactly since $A(\boldsymbol{r} - \boldsymbol{r}'') \neq \delta_{\boldsymbol{r}'',\boldsymbol{r}}$) the zeros of $\bar{F}$ in the diagonal elements ($\boldsymbol{r}'' = \boldsymbol{r}, \delta_{\boldsymbol{r}'',\boldsymbol{r}} = 1, A(\boldsymbol{r} - \boldsymbol{r}'') = 1$) given by

$$\omega - E_0 - U(\boldsymbol{r}) + i\delta \approx 0 \text{ at position } \boldsymbol{r} \tag{18}$$

Therefore, if $\boldsymbol{r}$ is on the barrier with height $U_0$, $U(\boldsymbol{r}) = U_0$ and the quasiparticle at $\boldsymbol{r}$ has energy

$$E(\boldsymbol{r}) \approx E_0 + U_0 \equiv E_b \text{ } (\boldsymbol{r} \text{ on the barrier).} \tag{19}$$

On the other hand, if $\boldsymbol{r}$ is more than the radius of $A(\boldsymbol{r})$ away from the barrier, we have $U(\boldsymbol{r}) = 0$ (i.e., no perturbation from the regular atomic potential) and

$$E(\boldsymbol{r}) = E_0 \text{ } (\boldsymbol{r} \text{ far away from the barrier).} \tag{20}$$

In the intermediate distance away from the barrier, $E(\boldsymbol{r})$ will assume an energy in between Eqs. (19) and (20).

The approximated equality in Eq. (19) would become exact if we had $A(\boldsymbol{r} - \boldsymbol{r}'') = \delta_{\boldsymbol{r}'',\boldsymbol{r}}$. However, for a realistic flat band structure, the finite width and the detailed realistic functional form of $A(\boldsymbol{r})$ introduces the following additional effects on the zeros of $\bar{F}$ considering its non-zero off-diagonal elements.

(i) Due to the finite width of $A(\boldsymbol{r})$, the effects of $U(\boldsymbol{r})$ is enhanced in Eq. 15 and we have $\Delta E_b (\equiv E_b - E_0)$ larger than $U_0$.

(ii) At the same time, considering that in reality the area of the flat band may be a small fraction of the 1st Brillouin zone, we have $A(\boldsymbol{r} = \boldsymbol{0}) < 1$ resulting in a reduction of $\Delta E_b$.

Therefore we may end up with the similarity of the barrier height $U_0$ and the elevation $\Delta E_b$ of the barrier-bound state energy in realistic materials as a result of the balance between these two competing factors dependent on the detailed flat band structure.

## Conclusion

Green function pole-analysis using real-space Dyson equation shows that counter-intuitive barrier-bound states are possible in flat-band systems. This does not restrict the orbital character of the barrier-bound states and it may be different from those of the flat band part of the band structure. The phenomenon is believed to be confirmed in a real system of Pd monolayer on a flat oxide substrate with Stoner magnetism using spin-polarized STM technique and through comparison with real-space QPI simulations using realistic 2D Pd band structure [4].

## Acknowledgement

The author would like to thank Steve Johnston and Peter Hirschfeld for useful discussion. The work is supported by National Research Foundation (NRF) through the Basic Science Research Programs (No. 2017R1D1A1B01016186) and the Pioneer Research Center Program (No. 2013M3C1A3064455).

## References

1. Nunner, T. S., Chen, W., Andersen, B. M., Melikyan, A. & Hirschfeld, P. J. "Fourier transform spectroscopy of d-wave quasiparticles in the presence of atomic scale pairing disorder," *Physical Review B* **73**, doi:10.1103/PhysRevB.73.104511 (2006).
2. This is the result of $\sum_{\boldsymbol{r}\in\mathcal{L}(A)} = \frac{1}{a^2 \sin\theta}\int_{\boldsymbol{r}\in A} d\boldsymbol{r}$ and $\frac{1}{L^2}\int_{\boldsymbol{r}\in A} d\boldsymbol{r} = \frac{1}{N^2}\sum_{\boldsymbol{r}\in\mathbb{N}(A)}$.
3. In the absence of the scattering potential, the density of states is uniform everywhere and $\rho(\boldsymbol{r},\omega) = -\frac{1}{\pi}\mathrm{Im}G(\boldsymbol{r},\boldsymbol{r},\omega) = -\frac{1}{\pi}\mathrm{Im}G_0(\boldsymbol{0},\omega) = -\frac{1}{\pi}\mathrm{Im}\sum_{\boldsymbol{k}\in\boldsymbol{1^{st}BZ}}^{M}\frac{1}{\omega+i\delta-\varepsilon_{\boldsymbol{k}}} = \sum_{\boldsymbol{k}\in\boldsymbol{1^{st}BZ}}^{M}\delta(\omega-\varepsilon_{\boldsymbol{k}})$
4. To be published.